\documentclass[useAMS,usenatbib]{mn2e}
\usepackage{placeins}
\usepackage{epsf}
\usepackage{amssymb}
\usepackage{amsmath}
\usepackage[usenames]{color}

\def\plotone#1{\centering \leavevmode
\epsfxsize=\columnwidth \epsfbox{#1}}
\def\plottwo#1#2{\centering \leavevmode
\epsfxsize=.99\columnwidth \epsfbox{#1} \hfil
\epsfxsize=.99\columnwidth \epsfbox{#2}}

\newcommand{\be}{\begin{equation}}
\newcommand{\ee}{\end{equation}}

\def\***#1{\textbf{\textsf{*** #1 ***}}}

\title[$z$ from SZ]{(No) dimming of X-ray clusters beyond $z\sim 1$ at
  fixed mass: crude redshifts and masses from raw X-ray and SZ data.}
\author[Churazov et al.]{E.~Churazov,$^{1,2}$ 
  A.~Vikhlinin,$^{3,2}$ R.~Sunyaev,$^{1,2}$
 \newauthor \\
$^1$ Max-Planck-Institut f\"ur Astrophysik, Karl-Schwarzschild-Strasse 1, 85741
Garching, Germany\\
$^2$ Space Research Institute (IKI), Profsoyuznaya 84/32, Moscow 117997, 
Russia\\
$^3$ Harvard-Smithsonian Center for Astrophysics, 60 Garden St.,
Cambridge, MA 02138, USA \\
}

\begin{document}

\pagerange{\pageref{firstpage}--\pageref{lastpage}}

\maketitle

\label{firstpage}
\begin{abstract}
Scaling relations in the $\Lambda$CDM Cosmology predict that for a
given mass the clusters formed at larger redshift are hotter,
denser and therefore more luminous in X-rays  than their local $z\sim 0$ counterparts. This effect overturns
the decrease in the observable X-ray flux so that it does not decrease at
$z > 1$, similar to the SZ signal. Provided
that scaling relations remain valid at larger redshifts, 
X-ray surveys will not miss massive clusters at any redshift,
no matter how far they are. \\ At the same time, the difference in scaling
with mass and distance of the observable SZ and X-ray signals from
galaxy clusters at redshifts $z\lesssim 2$ offers a possibility to
crudely estimate the redshift and the mass of a cluster. This might be especially useful for preselection of massive high-redshift clusters and planning of optical follow-up for overlapping surveys in X-ray (e.g., by SRG/eRosita) and SZ (e.g. Planck, SPT and ACT).
\end{abstract}


%

\sloppypar

\section{Introduction}

\begin{figure*}
\plottwo{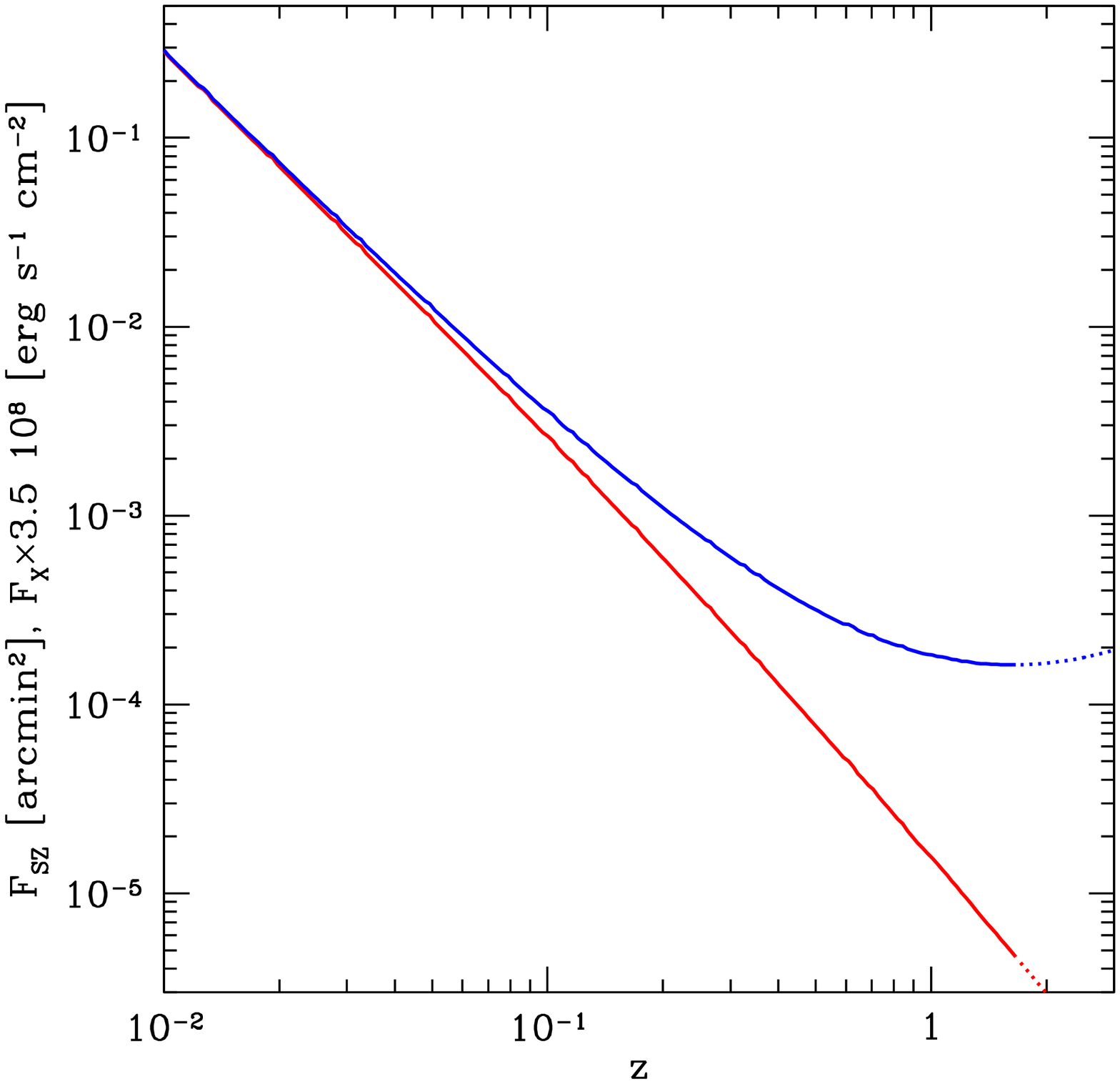}{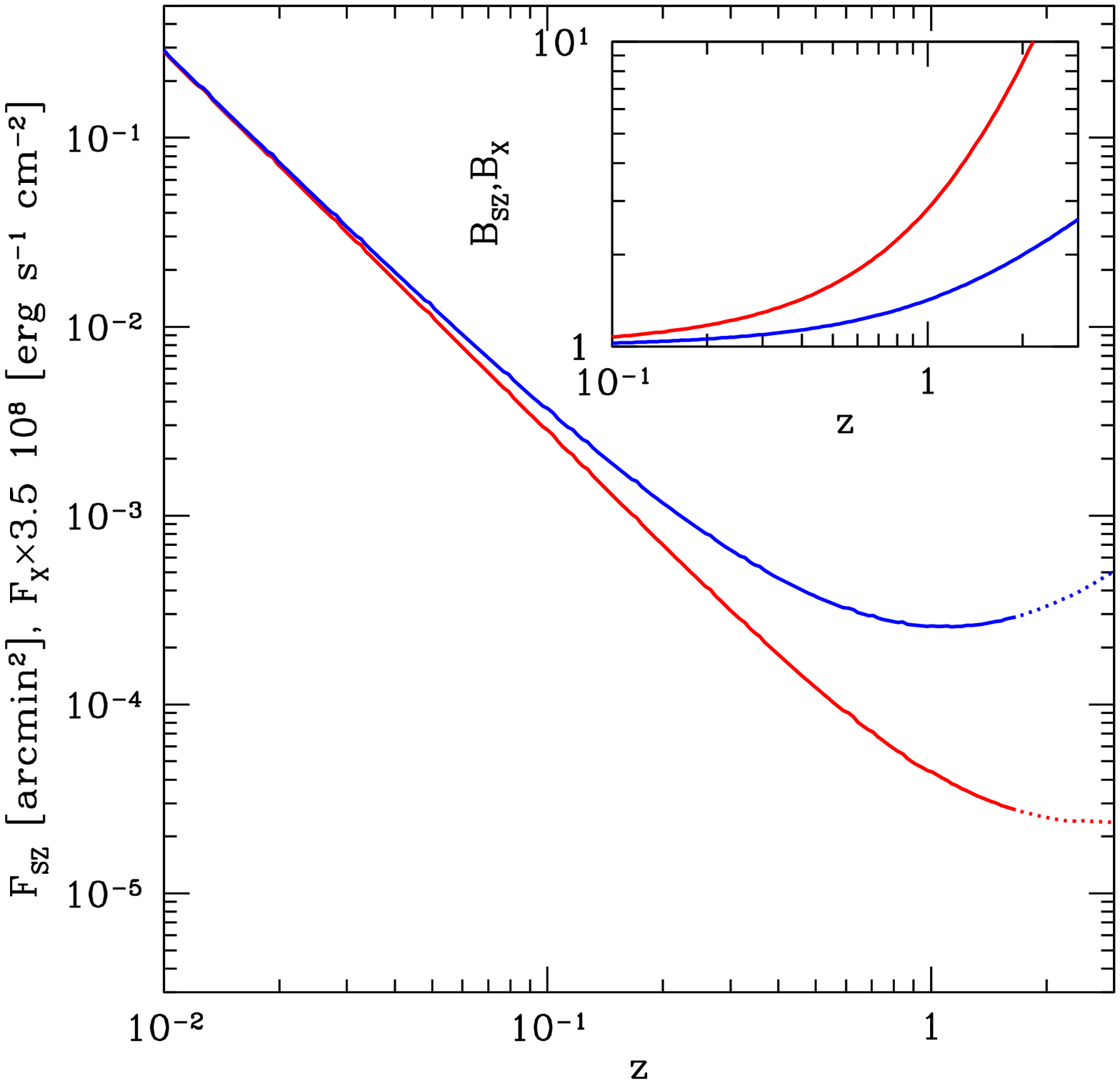}
\caption{{\bf Left:} Redshift dependence of the X-ray flux $F_{X}$ in
  the 0.5-2 keV band (red curve, in $\rm erg~s^{-1}~cm^{-2}$ multiplied by $3.5~10^{8}$) and SZ signal
  $F_{SZ}\equiv Y_{500}$ in units of $\rm arcmin^2$ for a cluster with a fixed mass
  $M_{500}=5~10^{14}~M_\odot$, temperature $T_X=5.5$~keV and size/luminosity.  At
  large redshifts the X-ray flux declines much more steeply than the
  SZ signal, which is just proportional to the square of the angular size of the cluster. {\bf Right:} The same plot when only mass is fixed at
  $M_{500}=5~10^{14}~M_\odot$, while other parameters follow
  empirical scaling relations (see eq. \ref{eq:escale}). Note that
  $F_X$ decreases up to $z\sim 1$ and then levels off. This means that
  if a cluster can be detected in X-ray survey at $z\sim 1$, then
  objects with similar mass {\it can be detected at any redshift}.
  The inset shows the boost factor in X-ray flux (red) and SZ signal (blue)
  calculated as the ratio of the the red curves in two panels. Clearly, the account for
  evolution (at fixed mass) boosts X-ray flux much stronger than the SZ signal. 
\label{fig:raw}
}
\end{figure*}

The abundance of massive clusters of galaxies is a sensitive probe of
the normalization of density perturbations power spectrum and the
Universe geometry \citep[e.g.,][]{1986MNRAS.222..323K}. In the
observable Universe there are about $10^5$ clusters with
masses\footnote{All masses quoted in the paper are at overdensity of
  500, unless stated otherwise} larger than $\sim
2~10^{14}~M_\odot$. Only a small fraction of these clusters is known
today, but the situation will change drastically with forthcoming
massive surveys in X-ray and microwave bands.

Main source of cluster candidates comes from X-ray
\citep[e.g.,][]{1998MNRAS.301..881E,2009ApJ...692.1033V} , SZ
\citep[e.g.,][]{2002ARA&A..40..643C,2014A&A...571A..29P} and optical
surveys \citep[e.g.,][]{2010ApJ...708..645R}. Each approach has its
own strengths and weaknesses. Finding clusters in X-rays is a very
powerful tool, capable of finding virialized cluster-scale objects at
great distances.  For example, in the upcoming SRG/eRosita all-sky
survey, the most distant cluster will be at $z\sim2$ with the mass
$M_{500}\sim 3~10^{14}~M_\odot$ \citep[e.g.,][]{2012arXiv1209.3114M}. This estimate assumes that one needs
$\sim 50$ counts to detect and classify an object as an extended
source and a cluster candidate. We argue below this limit in redshift
comes primarily from the lack of massive halos at higher redshifts,
rather then due to the dimming of X-ray signal with
distance. Therefore, with the upcoming SRG/eRosita data all massive
clusters in the observable Universe can be detected, no matter at what
redshift they are. The same is of course true for SZ-surveys (e.g., SPT or ACT).

Working with large number of candidates which are detected at the limit
of SRG/eRosita sensitivity is a challenge and requires a massive optical
followup program to measure redshifts. In many cases having even a crude
estimate of the redshift for distant clusters would be very helpful.

Recent progress in sensitivity and angular resolution of SZ facilities makes it possible to quickly detect clusters using the ground based instruments like CARMA or MUSTANG (exposures on the order of 20 minutes). However significant part of the new discovered clusters in X-Ray and microwave surveys will overlap. We expect the discovery of many tens of thousands of such objects for which both X-Ray and SZ fluxes will be available. It will be very difficult task to find observing time on big telescopes to measure redshifts (even photometric) for all these objects. Mentioned above  possibility of the rough redshift and mass estimate using just raw X-Ray and SZ-data might permit to separate immediately clusters which need small, medium or big telescopes for much more precise redshift measurements. This will help a lot to plan optical follow up. Situation obviously will be much easier on the regions of sky covered by extended optical redshift surveys like for example JPAS.

 Already in 1970 it
was recognized that different dependencies of surface brightness of
clusters in X-Rays and microwave bands on the gas density
$n_e$, temperature $T_e$ and the physical size of the cluster $L$,
namely, $I_X\propto n_e^2L$ (for X-rays) and $Y\propto n_e T_e L$ (for
microwave band), can be used to determine $L$ independently of its
redshift \citep[][page 16]{1970Ap&SS...7....3S}.  It was assumed that X-Ray data
provide us with the value of $T_e$. It is possible to compare X-ray and
CMB fluxes from clusters of galaxies and come to the same
conclusion \citep{1980ARA&A..18..537S}. Knowing $L$ and the angular
size of the cluster in X-Rays or in microwave band it is possible to
estimate the distance to the cluster. This permits to measure the
Hubble constant for any individual cluster if its redshift is known
\citep{1978ApJ...226L.103S,1978obco.meet....1G,1979A&A....75..322C,1979MNRAS.187..847B,2006ApJ...647...25B}. More than 1000 counts are needed to measure the gas temperature in clusters using observatories like Chandra or XMM-Newton. NUSTAR spacecraft, hard X-Ray grazing incidence telescopes aboard of Astro-H and SRG/ART-XC will open a way to measure electron temperature in most massive selected clusters with $kT_e > 10$ keV. This will give an additional information about evolution of the hot gas temperatures  inside clusters of galaxies as function of the redshift and will permit to take into account the relativistic corrections to SZ effect. However for faint objects the required exposure is long and the information on gas temperature will not be available for majority of objects in the SRG/eRosita survey. 

Cosmological evolution of  cluster properties with redshift and
 scaling relations found for real clusters add another aspect
 to the problem.  In particular, for clusters obeying observed scaling
relations (see Fig.~\ref{fig:raw}) the X-ray flux
(similarly to SZ signal) levels off at $z\sim 1$ and this minimal
level of X-ray flux can be used as a crude proxy to the mass of
distance clusters. At the same time the ratio of the X-ray and SZ
fluxes turns out to be weakly dependent on the mass (see
\citet{2012A&A...543A.102P} and \S\ref{sec:plane}) and can serve as a
crude proxy to the redshift. While these proxies suffer from the large
intrinsic scatter of cluster parameters, they might nevertheless be
useful for SRG/eRosita, SPT, ACT, which are planning to detect
more than 100 000 clusters on the whole sky. For these numerous (and mostly faint) objects a combination of X-ray and SZ
fluxes will provide a simple and convenient tool to approximately
position clusters in the mass-redshift diagram. This approach is similar to color-magnitude diagrams extensively used in infrared and optical astronomy.

For the purposes of this paper the relevant mass range is $M_{500}\sim
2~10^{14} - 2~10^{15} ~M_\odot$; the relevant redshift range is $\sim
0.1-2$.

Throughout the paper we adopt the following cosmological parameters:
$\Omega_M=0.27$, $\Omega_\Lambda=0.73$, $\Omega_bh^2=0.022$, $h=0.7$,
$w_0=-1$, $w_a=0$, $n=0.95$, $\sigma_8=0.79$. These parameters are
close to values derived/used in \citet{2009ApJ...692.1033V} from the
analysis of X-ray selected clusters at $0.02<z<0.9$. The choice of
this particular set is motivated by the use of the cluster scaling
relations based on the same data.

\section{X-ray and SZ signals and scaling relations}
\label{sec:scale}

\FloatBarrier

The total X-ray flux (in ${\rm erg~s^{-1}~cm^{-2}}$ in a given
energy band from $\epsilon_1$ to $\epsilon_2$ keV) from a galaxy cluster at redshift $z$ can be written as
\begin{eqnarray}
F_X\propto\frac{n_e^2 R^3}{D_L^2} C_X(T,z),
\end{eqnarray}
where $R$ is the physical size of the cluster, $n_e$ is the electron density,
$T$ is the gas temperature, $D_L$ is the luminosity distance to the
cluster. A factor $C_X(T,z)$ is the conversion coefficient from the
rest frame flux into observer frame for a given $T$ and $z$: 
\begin{eqnarray}
C_X(T,z)=\int_{\epsilon_1(1+z)}^{\epsilon_2(1+z)} S(\epsilon,T)d\epsilon,
\end{eqnarray}
where $S(\epsilon,T)$ in units of ${\rm erg~s^{-1}~cm^3}$ is the
emissivity at energy $\epsilon$
of an optically thin plasma with unit electron density. For hot clusters
($T>3$ keV) the emissivity is largely due to thermal
bremsstrahlung. In the arguments below we use the expression
\begin{eqnarray}
S(\epsilon,T)\propto \frac{e^{-\epsilon/kT}}{T^{1/2}}\approx \frac{1}{T^{1/2}},
\label{eq:b_simple}
\end{eqnarray}
which corresponds to the thermal bremsstrahlung spectrum for $\epsilon
\ll kT$. We note, however, that this expression is not accurate. First
of all, with account for the gaunt-factor the bremsstrahlung spectrum
(in the relevant temperature and energy ranges) is better described by
the expression $\displaystyle S(\epsilon,T)\underset{\sim}{\propto}
\epsilon^{-0.3}T^{-0.25}e^{-\epsilon/kT}$. Secondly, the contribution
of heavy elements (lines and recombination continuum) is important. Therefore for all plots 
the realistic approximation of $S(\epsilon,T)$ is used, calculated for APEC
model \citep{2012ApJ...756..128F} in XSPEC
\citep{1996ASPC..101...17A}. The main difference of these more
accurate approximations compared to eq.\ref{eq:b_simple} is a much weaker temperature dependence of the
emissivity in the soft X-ray band\footnote{Note that SRG/eRosita will have maximal sensitivity for clusters in the 0.5-2 keV band}. Nevertheless, for the sake
of simplicity of argument we use below eq.\ref{eq:b_simple}. With this
definition $\displaystyle
C_X(T,z)\underset{\sim}{\propto}(1+z)/\sqrt{T}$.

The total $Y_{SZ}$ parameter for the same cluster is
\begin{eqnarray}
F_{SZ}\equiv Y_{SZ}\propto \frac{n_e R^3}{D_A^2}
\sigma_T\frac{kT}{m_ec^2},
\end{eqnarray}
where $\sigma_T$ is the Thomson cross section, and $D_A$ is the
angular diameter distance. 

If $n_e$, $R$ and $T$ are fixed, then the changes of the $F_X$ and
$F_{SZ}$ are primarily driven by the factors $\displaystyle
\underset{\sim}{\propto} \frac{(1+z)}{D_L^2}$ and $\displaystyle
\frac{1}{D_A^2}$ respectively (Fig.\ref{fig:raw}, left panel). As
expected, the X-ray flux declines much faster than the SZ
signal. Indeed, the ratio of X-ray and SZ signals changes with $z$ as
\begin{eqnarray}
R_{XSZ}=\frac{F_X}{F_{SZ}}\propto \frac{D_A^2}{D_L^2}\frac{n_e
    C_X(T,z)}{T}=\frac{n_e
  C_X(T,z)}{(1+z)^4T}\underset{\sim}{\propto}\frac{1}{(1+z)^3}.
\label{eq:rxsz0}
\end{eqnarray}
However, the situation changes if instead of fixing all parameters, we
fix only the mass of a cluster and let the the other properties to
vary with redshift according to the cosmological scaling relations. In the
simplest scenario of a self-similar growth of halos the
characteristics of a virialized object, collapsing at redshift $z$ obey the
following scaling relations: $\displaystyle R \propto M^{1/3}E^{-2/3}$,
$\displaystyle T \propto \frac{GM}{R}$ and $\displaystyle n_e \propto
E^2$ \citep[e.g.][]{1986MNRAS.222..323K,1998ApJ...495...80B}, where $\displaystyle
E(z)=\frac{H(z)}{H(0)}=\sqrt{\Omega_\Lambda+\Omega_M(1+z)^3}$. 
Obviously, at a fixed mass the cluster collapsing at larger $z$ is
expected to be hotter, denser and more compact, boosting the X-ray
luminosity of the cluster. This boost in luminosity largely compensate for the $D_L^2$ factor in
$F_X$. Indeed,
\begin{eqnarray}
F_X\propto \frac{n_e^2 R^3 C_X(T,z)}{(1+z)^4 D_A^2}\propto
M\frac{E^2 C_X(T,z)}{(1+z)^4D_A^2}.
\end{eqnarray}
At $z\sim 1-2$, $D_A\approx {\rm const}$ and then declines as $\sim 1/z$, $E$
approximately scales as $(1+z)^{3/2}$ and $T\propto (1+z)$. Therefore
\begin{eqnarray}
  F_X \underset{\sim}{\propto}\frac{M}{D_A^2 (1+z)^{1/2}}.
\label{eq:fx_simple}
\end{eqnarray}
The redshift dependence of the $F_X$ curve changes dramatically
(Fig.\ref{fig:raw}, right panel\footnote{Note, that this plot
  corresponds to empirical scaling relations, rather than self-similar
ones. Nevertheless, the qualitative picture is the same.}), the steep decline is reverted and
overall behavior at $z\gtrsim 1$ resembles that of $F_{SZ}$. In other
words if a cluster can be detected in X-ray survey at $z\sim 1$, than
objects with similar mass {\it can be detected at any redshift},
assuming that the self-similar cluster evolution model is valid
  for extreme masses and very high redshifts. This
implies that with a sufficient sensitivity both SZ and X-ray surveys
have access to all massive clusters at all redshifts in the observable
Universe, even at redshifts where massive clusters are not expected to
exist.

In terms of the ratio of $F_X$ and $F_{SZ}$ at $z\gtrsim 1$
\begin{eqnarray}
R_{XSZ}=\frac{F_X}{F_{SZ}}\propto \frac{E^2 C_X(T,z)}{(1+z)^4E^{2/3}}
\underset{\sim}{\propto} \frac{1}{(1+z)^{3/2}},
\label{eq:rxsz1}
\end{eqnarray}
i.e. much weaker than for the case when all cluster properties are
fixed (c.f. eq. \ref{eq:rxsz0}). Similar expression was used by \citet{2012A&A...543A.102P} (see eq.~3 there) to verify possible redshift solutions for cluster candidates detected by Planck. 

The same conclusions can be formulated in terms of the X-ray surface
brightness at a given energy $\epsilon$ and Compton $y$-parameter
$\displaystyle y=\int \sigma_T n_e \frac{k T_e}{m_e c^2}dl$ :
\begin{eqnarray}
I_X(\epsilon)&\propto&\frac{n_e^2 R~S(\epsilon,T)}{(1+z)^3} \\
y&\propto& n_e R T. 
\end{eqnarray}
For a cluster obeying self-similar scaling both the X-ray surface
brightness and the $y$-parameter grow strongly with distance. The
ratio of these two quantities, of course, behaves similarly to the
ratio of total fluxes -- see eq. \ref{eq:rxsz1}. 

In reality, scaling relations, involving physics of baryons, deviate
to various degree from the self-similar relations \citep[see, e.g.,][ for a recent work]{2015MNRAS.446.2629E}. For a sample of
X-ray selected clusters at $0.02<z<0.9$ the following relations
have been derived \citep{2009ApJ...692.1033V}:
\begin{align}
  L_{X, 0.5-2} =&~ 1.2~10^{44}~{\rm erg~s^{-1}} \times \nonumber \\
  & \left(\frac{h}{0.72} \right )^{-0.39} \left
(\frac{M_{500}}{3.9~10^{14}M_\odot} \right )^{1.61} E^{1.85} \nonumber
\\
\label{eq:escale}
T_{X} =&~ 5~{\rm keV} \left
(\frac{M_{500}}{3.0~10^{14}~h^{-1}M_\odot} \right )^{0.65} E^{0.65}\\
Y_{X} =&~ 3~10^{14}~M_\odot {\rm keV} \left (
\frac{M_{500}}{5.8~10^{14}~h^{1/2}M_\odot} \right )^{5/3} E^{2/3}
\nonumber
\end{align}
where $L_X, 0.5-2$ is the rest frame cluster luminosity in the 0.5-2 keV
band, $Y_X=M_{g,500} T_X$ and $\displaystyle F_{SZ}=\eta Y_{X}$. Here $\eta=\frac{\sigma_T}{\mu m_p (n_e/n_t) m_ec^2 D_A^2}$, where $\mu\approx 0.61$ is the mean atomic weight of a gas particle, and $n_e/n_t\approx 0.52$ is the fraction of electrons in the total gas number density. 

The above relations can be used straightforwardly to predict the
dependence of $F_X$ and $F_{SZ}$ on $z$ for a fixed cluster mass, as
shown in Fig.\ref{fig:raw}, (right panel). The conclusion that $F_X$
does not decrease with redshift beyond $z\sim 1$ holds both for pure
self-similar and empirical scaling relations. It is a generic
consequence of self-similar evolution in the $\Lambda$CDM cosmological
model.

\begin{figure}
\plotone{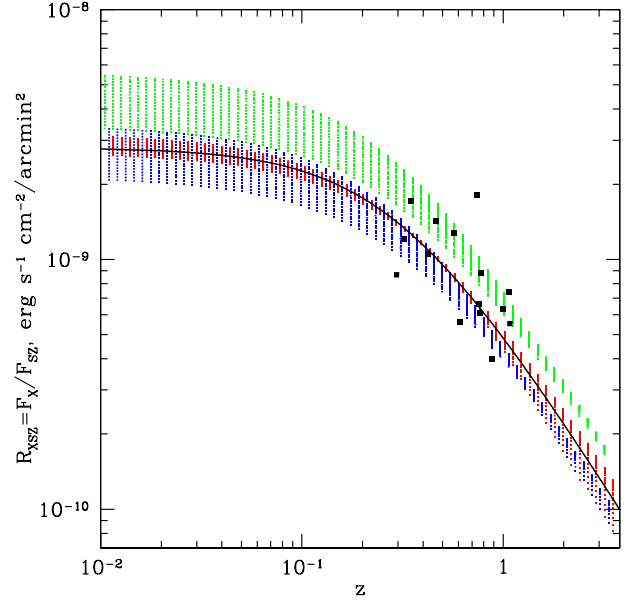}
\caption{Ratio $R_{XSZ}$ of the X-ray and SZ signals as a function of
  redshift. The shaded areas correspond to clusters in the mass range from $1~10^{14}$ to $2~10^{15}~M_\odot$ and the redshift range from 0.01 to 4 (note, that at large redshifts massive clusters are not expected to be present). The red area is for 
  clusters obeying scaling relations (\ref{eq:escale}) and the X-ray flux $F_X$ in the 0.5-2 keV band. The green area is for the same scaling relations, but uses $F_X$ in the 0.1-2.4 keV band. The blue area uses $Y_{500}$-$M$ relation from \citet{2011A&A...536A..11P} and the X-ray flux in the 0.5-2 keV band. Black squares show a small sample of SPT clusters with X-ray data from \citet{2011ApJ...738...48A}. 
  Below we adopt the $R_{XSZ}(z)$ dependence corresponding to red area as a base-line model, but the results can be readily reformulated for any other dependencies, shown in the figure.
 The black dashed line
  show a simple analytic approximation $\displaystyle
  R_{XSZ}=2.8~10^{-9}(1+(z/0.3)^{1.3})^{-1}$, which does not have physical meaning,
  but roughly reproduces the $R_{XSZ}$ dependence over the
  redshift range of interest.
\label{fig:ratio}
}
\end{figure}

\section{$F_{X}$ and $R_{XSZ}$ plane}
\label{sec:plane}
Given that for empirical relations $F_{X} \propto L_{X}\propto M^{1.61}$ and $F_{SZ}\propto
Y_{X} \propto M^{5/3}$ one can expect that in the ratio of these two
observables $\displaystyle R_{XSZ}=\frac{F_{X}}{F_{SZ}}$ the mass
dependence approximately cancels out and $R_{XSZ}$ is primarily a
function of $z$ \citep[see also][]{2012A&A...543A.102P}. This is further
illustrated in Fig.~\ref{fig:ratio}, which shows the value of
$R_{XSZ}$ for clusters with given mass
$M_{500}=10^{14}-2~10^{15}~M_\odot$ and slightly different definitions of $Y$ and $F_X$. Different colors correspond to: red - scaling relations (\ref{eq:escale}) and $F_X$ in the 0.5-2 keV band; green - the same scaling relations, but $F_X$ in the 0.1-2.4 keV band; blue - $F_X$ in the 0.5-2 keV band and $Y_{500}$-$M$ scaling relation from \citet{2011A&A...536A..11P}.  Although the masses differ by an
order of magnitude the $R_{XSZ}(z)$ curves are very similar. The black dashed line show a simple analytic
approximation of the data shown in red color:
\begin{eqnarray}
  R_{XSZ}\approx \frac{2.8~10^{-9}}{1+(z/0.3)^{1.3}}~{\rm erg~s^{-1}~cm^{-2}~arcmin^{-2}},
  \label{eq:rxsz}
\end{eqnarray}
which does not have physical meaning, but roughly reproduces the
expected $R_{XSZ}$ dependence over the redshift range of interest.
Black squares show a small sample of SPT clusters with X-ray data from \citet{2011ApJ...738...48A}. 
For the relevant redshift range (up to $z\sim 2$) the value of $R_{XSZ}$ changes by an
order of magnitude, implying that it can be used to approximately
divide clusters in few redshift bins over $0<z<2$, despite the substantial
scatter of cluster properties around the mean scaling relations.

\begin{figure}
\plotone{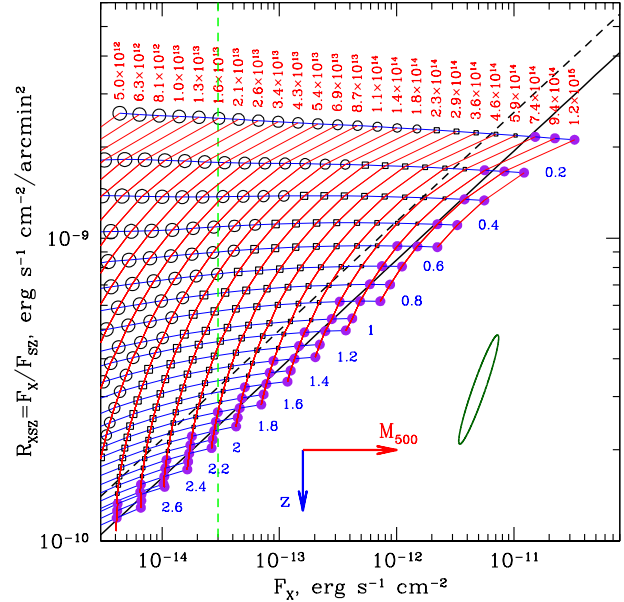}
\caption{Expected number of clusters with a given
  $(F_X,R_{XSZ})$. Clusters with $T<3$ keV are marked with circles,
  while hotter clusters are marked with squares. The size of the
  symbol is proportional to the log of the number of clusters per grid
  cell, i.e.  $\displaystyle \frac{dN}{dMdz}\times \Delta z \Delta
  M$. Red and blue lines correspond to the constant mass and redshift
  respectively.  Purple circles mark 20 most massive clusters in each
  redshift bin. The approximate orthogonality of the constant redshift
  and constant mass curves suggest that a combination of X-ray and SZ
  data is a useful indicator of the object redshift and mass. The
  green dashed line is the cluster detection threshold for eRosita
  all-sky survey. Dark green ellipse shows typical uncertainty in the
  position of a cluster in the $(F_X,R_{XSZ})$ plane due to the
  intrinsic scatter in $F_X$ (dominant component of the scatter) and
  $Y$. Without the scatter, the expected number of (most massive)
  clusters below black solid line is $\sim$10 per each $\left [F_X;
    2\times F_X \right ]$ bin. To compensate for the intrinsic scatter from the mean scaling relations one has to use less strict selection cut (dashed line) to provide 90\%
  completeness of the sample for these massive clusters.
\label{fig:map}
}
\end{figure}

We now consider
how clusters populate the $\displaystyle ( F_{X}$,
$R_{XSZ})$ plane, using predictions of the
abundance of virialized halos at a given redshift/mass. To this end, we
use the mass function of \citet{2008ApJ...688..709T}. We start by
calculating cluster abundance $\displaystyle \frac{dN}{dMdz}$ for a
grid over $z$ and $M=M_{500}$. For
each pair of $z$ and $M$ we use scaling relations (\ref{eq:escale}) to
calculate $F_X$ and $F_{SZ}$. The results of these calculations are
shown in Fig.~\ref{fig:map}. The size of the circle is proportional to
the log of the number of clusters per grid cell, i.e.  $\displaystyle
\frac{dN}{dMdz}\times \Delta z \Delta M$. Only cells with more than
one cluster are shown. Cells, corresponding to the constant mass or
redshift are connected with red and blue lines respectively.

As expected, blue lines are almost horizontal, implying that the ratio
$R_{XSZ}=\frac{F_{X}}{F_{SZ}}$ changes with $z$, but is weakly
dependent on mass (see also Fig.~\ref{fig:ratio}). Therefore, $R_{XSZ}$ can serve as a redshift indicator, at least for clusters more massive than
$\sim 2~10^{14}~M_\odot$.  At the same time the red lines
(constant mass) are nearly vertical at $z\gtrsim 1$ and are slightly
inclined at lower redshifts. This behavior can be traced back to Fig.\ref{fig:raw} (right panel) which shows that at fixed mass the $F_X$ curve levels off at $z\sim 1$. The curves, corresponding to clusters with different masses will have different minimum flux, which is proportional to the mass of the cluster, if one adopts simplified $F_X-M$ relation according to eq.\ref{eq:fx_simple}. With the scaling relations adopted here (eq.~\ref{eq:escale}) the mass dependence is steeper (see below).  

The approximate orthogonality of the
constant redshift and constant mass curves suggests that a combination
of X-ray and SZ data is a useful indicator of the object redshift and
mass. In other words, there is an approximate mapping of $F_X$ to
$M_{500}$ and $R_{XSZ}$ to $z$, which is most straightforward in the
interesting range of $z\sim$1-2.

\begin{figure}
\plotone{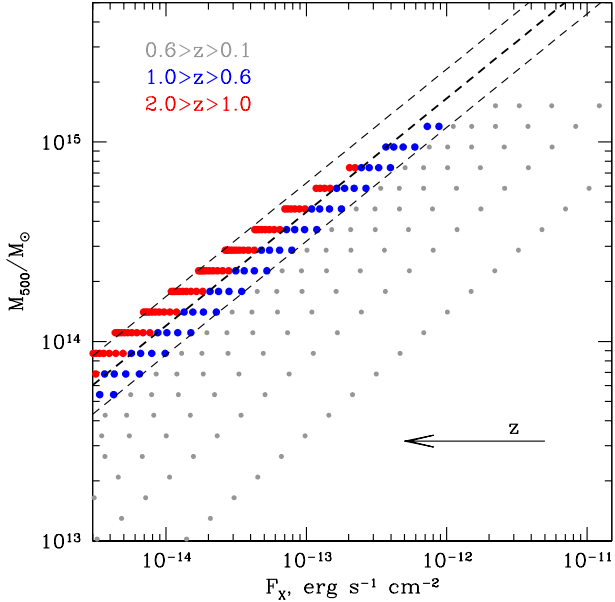}
\caption{Expected correlation of the observed 0.5-2 keV X-ray flux and the mass of the cluster for groups of objects at different redshift. Redshift changes from right to left in steps of 0.1. For nearby clusters ($z<0.6$, grey circles) the spread in the $F_X$ at a given mass is very large. On the contrary, for $z>0.6$ (blue and red points) there is relatively tight correlation between the X-ray flux and the mass. The tightness of the correlation is due to evolution of cluster properties with redshift. The thick black dashed line shows a simple analytic approximation of this correlation $\displaystyle M_{500}=1.2~10^{14} \left (  \frac{F_X}{10^{-14}} \right )^{0.57}$. Thin dashed lines show the same line scaled up and down by a factor 1.4.
\label{fig:fx_mass}
}
\end{figure}

The expected correlation of the observed X-ray flux and the cluster
mass is further illustrated in Fig.~\ref{fig:fx_mass}. The same
mass/redshift bins as in Fig.~\ref{fig:map} are used, but this time the mass is
plotted as a function of X-ray flux.  Clusters from different redshift
groups are marked with different colors. For nearby clusters ($z<0.6$,
black circles) the spread in $F_X$ at a fixed mass is huge. On the contrary, for $z>0.6$ (blue and red points) there is a good correlation between the X-ray flux and the mass. This
correlation appears due to evolution of cluster properties with
redshift. The black dashed line shows a simple analytic approximation
of this correlation
\begin{eqnarray}
M_{500}=1.2~10^{14} \left ( \frac{F_X}{10^{-14}} \right )^{0.57}
  \label{eq:fx_mass}.
\end{eqnarray}

\begin{figure}
\plotone{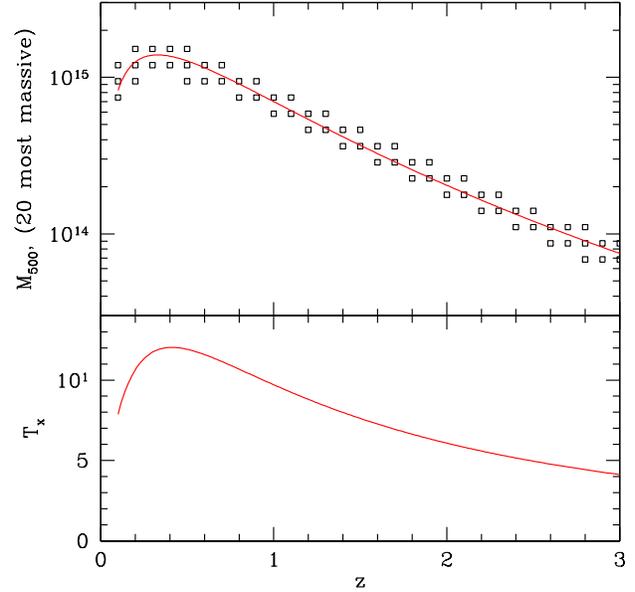}
\caption{Most massive 20 clusters in each redshift bin according the
  mass-function of \citet{2008ApJ...688..709T}. The decrease of the mass at small $z$  is due to decrease of the volume.  The red line in the
  top panel show a simple analytic approximation of the redshift
  dependence $M_{20}(z)$ of typical masses of such clusters. The bottom panel shows the
  expected gas temperature, calculate using empirical relation (eq.\ref{eq:escale})
for $M_{20}(z)$. Up to $z\sim3$ all these clusters have $T_X \gtrsim
5$ keV.
\label{fig:mz}
}
\end{figure}

The most stringent limits on the growth of structure come from the
most massive clusters at each redshift. 
Fig.~\ref{fig:mz} shows the
mass and temperature dependence for the 20 most massive clusters in each redshift bin according the
  mass-function of \citet{2008ApJ...688..709T}. The characteristic mass
changes by an order of magnitude (red line in the top panel of
Fig.~\ref{fig:mz}), while the characteristic temperature stays above
4-5 keV over the redshift range 0-3. The relativistic corrections to the SZ signal can be important for most massive clusters at $z\sim0.5$, while for more distant ($z\gtrsim 1$) massive clusters they are expected to be small. The locus of these massive
  clusters in the $(F_X,R_{XSZ})$ plane is shown with purple
  circles. It is clear that these clusters form a elongated and narrow
  band in the plane, implying that picking clusters candidates from
  this band will provide a useful sub-sample for further studies. With out definition of X-ray and SZ signals the most massive clusters at a given $F_X$ correspond to those below the line $\displaystyle R_{XSZ,10}=3.7~10^{-10} \left ( \frac{F_X}{10^{-13}{~\rm erg ~s^{1}~cm^{-2}}}\right )^{-0.36}$ (black solid line in Fig.~\ref{fig:map}). There are $\sim$10 massive clusters below this line per each X-ray flux interval $\left [F_X; 2\times F_X \right ]$ over the whole range of interest.
  
The above discussion completely neglects the scatter of cluster
properties around the adopted scaling relations. This scatter in
$L_{x}$ is considerable
\citep[e.g.,][]{2006ApJ...648..956S,2009ApJ...692.1033V} and will
spread the clusters in the $(F_X,R_{XSZ})$ plane. We assume that the scatter in $\log F_X $ and $\log Y $ is 0.4 and 0.13 respectively \citep{2009ApJ...692.1033V} and that the scatter is independent in these two variables. With this assumptions the expected level of the scatter in the $(F_X,R_{XSZ})$ plane is shown with a green ellipse in Fig.~\ref{fig:map} with $\log$ of the major and minor axes of 0.57 and 0.09 respectively. Note, that since the scatter in $F_X$ is substantially larger than the scatter in $Y$, we choose to plot an ellipse, corresponding to 68\% probability for one degree of freedom (rather than for two). When scatter is included the selection cut for massive clusters has to be modified. Assuming that we want 90\% of the most massive clusters to be captured by the selection (i.e. those objects which without intrinsic scatter would be below the solid line), we need to increase the selection cut by as factor 1.34 (black dashed line in Fig.~\ref{fig:map}), as demonstrated by Monte Carlo simulations. Thus, selecting objects with $R_{XSZ}<1.34~R_{XSZ,10}$ will ensure that 9 out of 10 most massive clusters will not be lost from the sample due to intrinsic scatter. Of course this relaxed selection will contain factor $\sim 10^2$ more objects than our target sample at $F_X\sim 3~10^{-14} {~\rm erg ~s^{-1}~cm^{-2}}$. I.e., about $10^3$ cluster candidates has to be followed up in order to identify 10 most massive ones. This is nevertheless a substantial improvement compared to the total number of objects per $F_X$ bin without $R_{XSZ}$ cut, which is another factor of $\sim 10^2$ larger.

We also used a  Monte Carlo approach to estimate the uncertainties in evaluating $z$ via eq.(\ref{eq:rxsz}). For each redshift the scatter is added to $F_X$ and $F_{SZ}$, the value of $R_{XSZ}$ is calculated and eq.(\ref{eq:rxsz}) is used to recover the redshift.  The estimated uncertainties in $z$ correspond  to lower/upper 17\% of the probability distribution. For $z=1,1.5$ and 2, these intervals are: $1.0^{+0.42}_{-0.32}$,  $1.5^{+0.61}_{-0.45}$ and $2.0^{+0.77}_{-0.57}$. As expected, the uncertainties are large, but suitable for pre-selection of massive and distant clusters for further studies.

For the mass determination using eq.(\ref{eq:fx_mass}) the uncertainty is factor $\sim$1.4 in the normalization of eq.(\ref{eq:fx_mass})  (see previous section). Additional uncertainty due to the scatter in $F_X$ adds another $\sim$22\% to the mass estimate.

Finally, we note that $(F_X,R_{XSZ})$ is not the only pair of variables which can be used
for preselection of clusters, but it is one of the most
straightforward and simple combinations of the observables. In practice the selection strategy has to be optimized for a particular set of instruments/observables used in the X-ray and SZ surveys.


\section{Conclusions}
In $\Lambda$CDM Universe distant clusters with a given mass are denser
and hotter compared to their local counterparts. This boosts strongly
the rest frame X-ray luminosity of distant clusters and compensates
for the dimming of the X-ray signal with distance. This implies that
similarly to SZ surveys, X-ray telescopes are able to detect massive
cluster over entire observable Universe. If the cluster can be
detected at $z\sim 1$, similar mass objects can be detected at any
redshift. 

A combination of X-ray and SZ observables happens to be a simple
indicator of $z$ and $M$ (see eq.\ref{eq:rxsz} and \ref{eq:fx_mass} and Fig.\ref{fig:ratio} and \ref{fig:fx_mass}), which can be used to approximately classify
clusters in broad mass/redshift bins for the spectroscopic followup. Provided
that local scaling relations hold for larger redshift the ratio
$R_{XSZ}$ of
X-ray flux to the total $Y$ is almost independent on the mass and
serves as convenient redshift indicator, while the X-ray flux itself
can be used as the mass proxy (for objects with sufficiently large redshift $z\gtrsim0.5$, or, equivalently, with sufficiently low $R_{XSZ}$).

\section{Acknowledgments}
We are grateful to the referee for a number of helpful comments and suggestions.
We acknowledge partial support by 
grant No. 14-22-00271 from the Russian Scientific Foundation.

\label{lastpage}
\end{document}